\documentclass[aps,prb,twocolumn]{revtex4}
\usepackage[ansinew]{inputenc}
\usepackage[pdftex]{graphics}
\usepackage{appendix}
\usepackage{graphicx}
\usepackage{amsmath}
\usepackage{amsthm}
\usepackage{bm}
\usepackage{layout}
\usepackage{float}
\usepackage{txfonts}
\usepackage{amsfonts}
\usepackage{amssymb}
\usepackage{dsfont}
\usepackage{hyphenat}

\usepackage[english]{babel}
\usepackage{latexsym}
\usepackage{amscd}
\usepackage{amsgen,amstext,amsbsy,amsopn}
\usepackage{math rsfs}
\usepackage{epsfig}

\usepackage{amssymb}
\usepackage{amsmath}
\usepackage{amsfonts}
\usepackage{color}
\def\ket#1{|#1\rangle}
\def\bra#1{\langle#1|}

\def\bp{\mathrm{p}}

\def\Tr{\mathrm{Tr}}

\newcommand{\Ident}{\mathds{1}}

\begin{document}
\title{Quantum Discord as Optimal Resource for Quantum Communication}
\author{Borivoje Daki{\'c}$^{1\dagger}$, Yannick Ole Lipp$^{1\dagger}$, Xiaosong Ma$^{2,3\dagger}$, Martin Ringbauer$^{1\dagger}$, Sebastian Kropatschek$^{3}$, Stefanie Barz$^{2}$, Tomasz Paterek$^{4}$, Vlatko Vedral$^{4,5}$, Anton Zeilinger$^{2,3}$, {\v C}aslav Brukner$^{1,3}$ \& Philip Walther$^{1,3}$}
\affiliation{
$^1$ Faculty of Physics, University of Vienna, Boltzmanngasse 5, A-1090 Vienna, Austria.\\
$^2$ Vienna Center for Quantum Science and Technology, Faculty of Physics, University of Vienna, Boltzmanngasse 5, A-1090 Vienna, Austria.\\
$^3$ Institute for Quantum Optics and Quantum Information, Austrian Academy of Sciences, Boltzmanngasse 3, A-1090 Vienna, Austria.\\
$^4$ Centre for Quantum Technologies, National University of Singapore, Block S15, 3 Science Drive 2, 117543 Singapore.\\
$^5$ Department of Atomic and Laser Physics, University of Oxford, Oxford OX1 3PU, UK.\\
$^\dagger$~ These authors contributed equally to this work (listed alphabetically).}

\begin{abstract}
Quantum entanglement is widely recognized as one of the key resources for the advantages of quantum information processing, including universal quantum computation~\cite{Nielsen&Chuang}, reduction of communication complexity~\cite{Buhrman2010,Brukner2002} or secret key distribution~\cite{Ekert1991}. However, computational models have been discovered, which consume very little or no entanglement and still can efficiently solve certain problems thought to be classically intractable~\cite{Knill,Meyer}. The existence of these models suggests that separable or weakly entangled states could be extremely useful tools for quantum information processing as they are much easier to prepare and control even in dissipative environments. It has been proposed that a requirement for useful quantum states is the generation of so-called quantum discord~\cite{Zurek,Henderson}, a measure of non-classical correlations that includes entanglement as a subset. Although a link between quantum discord and few quantum information tasks has been studied, its role in computation speed-up is still open and its operational interpretation remains restricted to only few somewhat contrived situations~\cite{Caves2,Cavalcanti,Datta,Horodecki2008}. Here we show that quantum discord is the optimal resource for the remote quantum state preparation~\cite{Bennett2001}, a variant of the quantum teleportation protocol~\cite{Bennett1993}. Using photonic quantum systems, we explicitly show that the geometric measure of quantum discord~\cite{Dakic} is related to the fidelity of this task, which provides an operational meaning. Moreover, we demonstrate that separable states with non-zero quantum discord can outperform entangled states. Therefore, the role of quantum discord might provide fundamental insights for resource-efficient quantum information processing.

\end{abstract}

\date{\today}
\maketitle

\textbf{\emph{Introduction.--}} Quantum computation and quantum communication is believed to allow for information processing with an efficiency that cannot be achieved by any classical device. It is usually assumed that a key resource for this enhanced performance is quantum entanglement~\cite{Schroedinger1935}. The creation and manipulation of entanglement, however, is a very demanding task, as it requires extremely precise quantum control and isolation from the environment. Thus, current experimental achievements are limited to rather small scale entangled systems \cite{Monz2011,Yao2011,Huang2011}. On the other hand there is no proof that quantum entanglement is necessary for quantum information processing (QIP) that can outperform its classical counterpart. The investigation of QIP protocols that allow for significant enhancements in the efficiency of data processing by only using separable states is of high interest. Obviously, such states have the benefit of being easier to prepare and more robust against losses and experimental imperfections. In fact, there are quantum computational models based on mixed, separable states, most notably the so-called deterministic quantum computation with one qubit (DQC1)~\cite{Knill}, which has recently been demonstrated experimentally~\cite{Ryan2005,White,Laflamme}. In this context, quantum discord has been proposed as the resource that can provide the enhancement for the computation~\cite{Caves1,Schack}, but its relation to the computational speed-up remains ambiguous~\cite{Dakic,Acin}. A relation to quantum communication was shown to exist only in few particular cases, for example in local broadcasting~\cite{Horodecki2008} and quantum state merging~\cite{Cavalcanti,Datta}.

Here we identify quantum discord as the crucial resource for a fundamental quantum information protocol, the remote state preparation (RSP). This protocol is a variant of quantum state teleportation in which the sender (Alice) knows the quantum state to be communicated to the receiver (Bob). The experimental implementation was performed on a quantum optical platform using polarization-correlated single photons. We find that for a broad class of states the fidelity of RSP is directly given by the geometric measure of quantum discord. This provides an operational meaning to this measure of ``quantumness'' of correlations in quantum information. Remarkably, states with no entanglement, yet non-zero geometric quantum discord can outperform entangled states in accomplishing RSP.

\textbf{\emph{Theory.--}} Two systems are correlated if together they contain more information than taken separately.
This intuitive definition is formally captured by (quantum) mutual information~\cite{Winter} $I({A:B})=H(A)+H(B)-H(A,B)$, where $A$ and $B$ are random variables. In classical probability theory $H(\cdot)$ stands for the Shannon entropy $H(\bp)=-\sum_ip_i\log p_i$, where $\bp=(p_1,p_2,\dots)$ is the probability distribution vector, while in the quantum case it denotes the von Neumann entropy $H(\rho)=-\mathrm{Tr}\rho\log\rho$ of a density matrix $\rho$. Classically, we can use the Bayes rule and find an equivalent expression for the mutual information $I(A:B)=H(A)-H(A|B)$, where $H(A|B)$ is the Shannon entropy of $A$ conditioned on the measurement outcome of $B$. For quantum systems, these two expressions are inequivalent and their difference defines a non-negative quantity, the so-called quantum discord~\cite{Zurek,Henderson}.

Whenever quantum discord vanishes the systems are classically correlated. A very simple expression for zero-discord states was proposed~\cite{DattaPhD}, which will be illustrated in the following example. Consider the case of a two-qubit state shared by Alice and Bob, where the Hilbert space of each qubit is spanned by the orthogonal states $\ket{0}$ and $\ket{1}$. A general zero-discord state $\chi$ can be written as
\begin{equation}
\chi=p_1\ket{0}\bra{0}\otimes\rho_1+p_2\ket{1}\bra{1}\otimes\rho_2,
\label{eqn:ClassicalState}
\end{equation}
where $p_1+p_2=1$ and $\rho_{1,2}$ are arbitrary states of the second qubit. Intuitively, this can be understood as a joint system containing one classical bit (cbit) and one quantum bit (qubit), $1\,\mathrm{cbit}\times1\,\mathrm{qubit}$. One of the systems can be identified as a cbit, because it is always in one of the perfectly distinguishable states $\ket{0}$ and $\ket{1}$. All other states possess some genuine quantum correlations.

\begin{figure}
\includegraphics[width=0.4\textwidth]{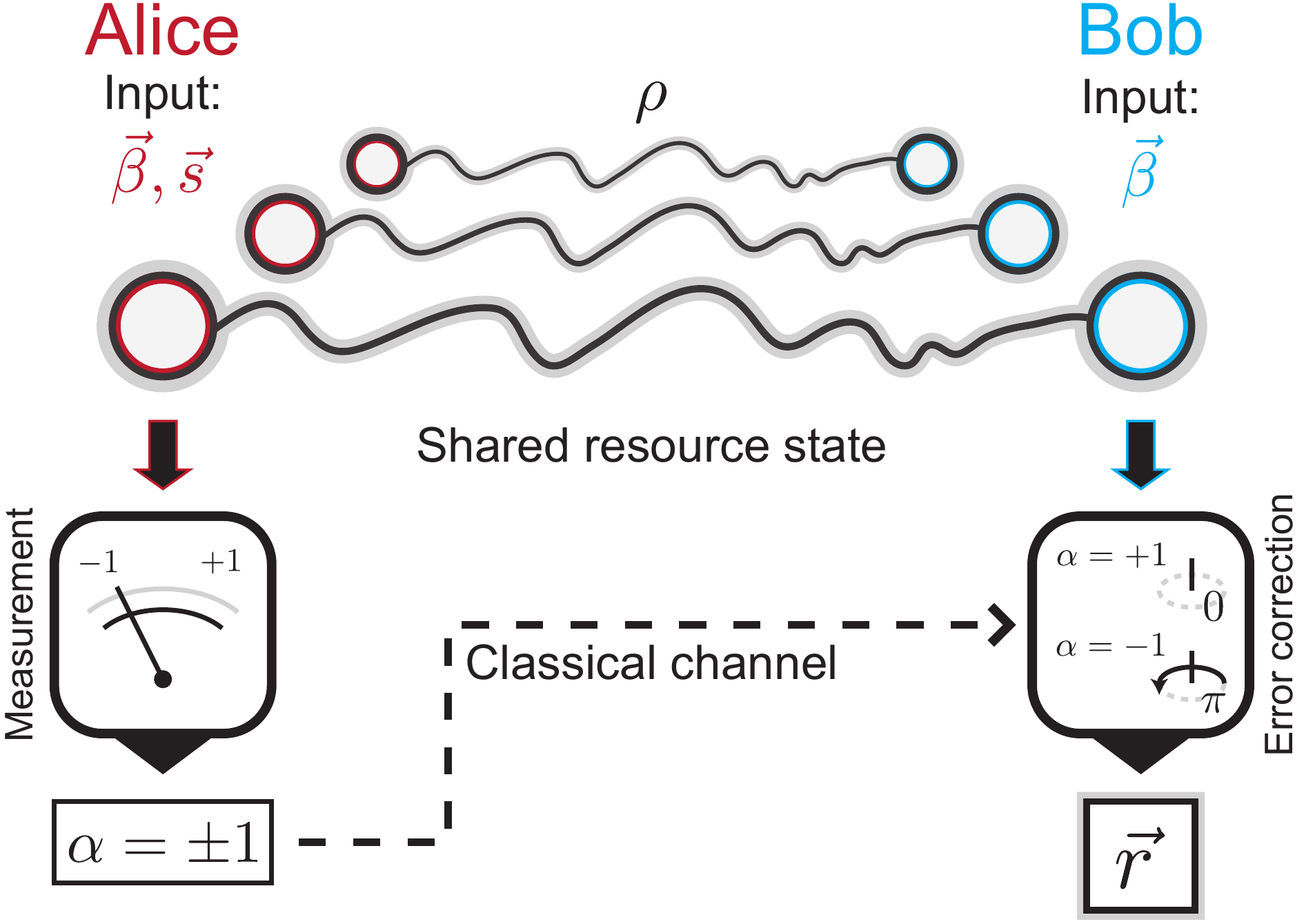}
\caption{Remote state preparation (RSP). Alice and Bob share a quantum state $\rho$. She is supposed to prepare a state $\vec{s}$ in the plane orthogonal to the direction $\vec{\beta}$ (that is announced to Bob as well). After performing the measurement along the direction $\vec \alpha$ she sends her result $\alpha=\pm1$ to Bob via a classical channel. Conditioned on the value of $\alpha$ Bob applies a correction to his qubit to obtain the state $\vec r$.}
\label{fig:RSPProtocol}
\end{figure}

Consider the representation of a two-qubit state $\rho$ in terms of local Pauli matrices $\{\sigma_1,\sigma_2, \sigma_3\}$,
\begin{equation}
\rho = \frac{1}{4} \left(\Ident \otimes \Ident + \sum_{k=1}^3 a_k \sigma_k \otimes \Ident + \sum_{l=1}^3 b_l \Ident \otimes \sigma_l + \sum_{k,l=1}^3 E_{kl} \sigma_k \otimes \sigma_l \right),
\end{equation}
where ${E_{kl} = \Tr(\sigma_k \otimes \sigma_l \rho)}$ are the elements of correlation tensor $E$. The vector $\vec a = (a_1,a_2,a_3)$ with components $a_k = \Tr(\sigma_k\rho_A)$ is the Bloch vector of the reduced density operator $\rho_A$ of Alice and similarly $\vec b$ for Bob. One can always choose a local reference frame on Alice's and Bob's sides such that the correlation tensor becomes a diagonal matrix in the Schmidt canonical form~\cite{Horodecki1996} $E=\mathrm{diag}[E_1,E_2,E_2]$, where $E_i^2$ are the eigenvalues of $E^{\mathrm{T}}E$. A zero discord state has the Schmidt form $E=\mathrm{diag}[E_1,0,0]$, which corresponds to correlation in one basis only.

There are many ways to quantify quantum correlations~\cite{Modi2011}. The advantage of the recently introduced geometric measure of quantum discord is that it can be evaluated explicitly and leads to an analytical closed form in many interesting cases~\cite{Dakic,Luo}. It is defined as the normalized trace distance to the set of classical states
\begin{equation}
\mathcal{D}^2 (\rho) \equiv 2 \min_{\chi} \left\|\rho - \chi\right\|^2 \equiv 2 \min_{\chi} \Tr(\rho - \chi)^2,
\label{eqn:GMD}
\end{equation}
where the minimum is taken over the set of zero-discord states $\chi$, i.e.\ states of the form of Eq.~\eqref{eqn:ClassicalState}. For states with maximally mixed marginals ($\vec a = \vec b = 0$) the geometric discord takes the following simple form~\cite{Dakic}
\begin{equation}
\mathcal{D}^2 (\rho)=\frac{1}{2}(E_2^2+E_3^2) ,
\label{eqn:GMDspecial}
\end{equation}
where $E^2_{2,3}$ are the two lowest eigenvalues of $E^{\mathrm{T}}E$. In fact, the same expression holds for a much larger class of states (see Methods for details). We will show that Eq.~\eqref{eqn:GMDspecial} captures the quality of RSP, thereby providing an operational meaning for geometric discord.

For the implementation of the RSP protocol Alice and Bob share a quantum state $\rho$, which can possess various correlations, e.g.\ classical correlation, entanglement and quantum discord. In the case of RSP, the task of preparing a specific state at Bob's location can be accomplished with fewer resources than in the case of quantum teleportation~\cite{Bennett1993}. Preparing an arbitrary unknown state via teleportation requires communication of two cbits, while sharing a maximally entangled state. However, if Alice only wants to remotely prepare a quantum state $\ket{\psi}$ on the equatorial plane of the Bloch sphere, e.g.\ $\ket{\psi}=(\ket{0}+e^{i\phi}\ket{1})/\sqrt{2}$, one cbit of communication is sufficient. The protocol that achieves this uses a maximally entangled state as a shared resource and a single cbit, which is Alice's measurement outcome along the direction of the state she wants to prepare. Depending on the value of the received cbit Bob needs to correct his qubit by a $\pi$ rotation about the $z$ axis to generate the state envisaged by Alice.

In the ideal case a shared maximally entangled state enables Alice to prepare deterministically any state in the equatorial plane of Bob's Bloch sphere. In general, when a mixed state is used as quantum resource for the RSP protocol, then Bob obtains a quantum state with a reduced fidelity~\cite{Bennett2005}. For the investigation of the underlying resource for the RSP protocol photon pairs with different polarization correlations are generated and shared by Alice and Bob. Alice uses this shared quantum state $\rho$ to remotely prepare a state $\vec s$ in the plane orthogonal to the direction $\vec{\beta}$ on Bob's side. She initializes the state preparation on Bob's side by performing a local measurement along the direction $\vec \alpha$. The measurement outcome $\alpha=\pm 1$ is then sent to Bob as one cbit of information. For $\alpha=-1$ Bob applies a $\pi$ rotation about $\vec\beta$ to his system, while no correction is required for $\alpha=1$. The resulting state on Bob's side is denoted by $\vec r$ (see Figure~1). To evaluate the efficiency of the protocol we first define a payoff-function $\mathcal{P}\equiv(\vec r \cdot \vec s)^2$, which is calculated for each run. Alice can optimize the payoff for a given $\vec s$ and $\vec\beta$ by her choice of the local measurement direction $\vec\alpha$. The fidelity of the protocol is calculated as an averaged payoff and minimized over all possible choices of the direction $\vec \beta$, which captures the worst case scenario. The payoff $\mathcal{P}$ is non-zero if and only if Alice is able to prepare a state effectively different from a completely mixed state. In particular $\mathcal{P}=1$, when using a maximally entangled state as shared resource, while $\mathcal{P}=0$ for a totally mixed state.

Careful analysis shows that the described RSP-fidelity is given by the following expression (see Methods for details)
\begin{equation}
\mathcal{F}\equiv \min_\beta \langle \mathcal{P}_{opt}\rangle = \frac{1}{2}\left( E_2^2 + E_3^2\right) ,
\label{eqn:RSPFidelity}
\end{equation}
where $E_2^2$ and $E_3^2$ are the two lowest eigenvalues of $E^\mathrm{T} E$. As noted above, this quantity captures the suitability of a certain resource state for RSP. We further observe that for a broad class of states the geometric measure of quantum discord has the form of Eq.~\eqref{eqn:GMDspecial} and therefore matches exactly the RSP-fidelity (see Methods for details). Apparently, in this case geometric discord is provided with an operational meaning. From Eq.~\eqref{eqn:RSPFidelity} it follows that $\mathcal{F}>0$ if and only if the correlation tensor has at least two non-zero eigenvalues. Recall that a zero-discord state has a correlation tensor of the form $E=\operatorname{diag}[E_1,0,0]$. In other words, $\mathcal{F}=0$ for exactly the states of the form of Eq.~\eqref{eqn:ClassicalState}. Therefore non-zero fidelity necessarily implies the presence of non-zero quantum discord.

We now consider the Werner states~\cite{Werner1989}, $\rho_W$:
\begin{equation}
\rho_W = \lambda\ \ket{\psi^-}\bra{\psi^-}+\frac{(1-\lambda)}{4}\Ident_4 ,
\label{eqn:Werner}
\end{equation}
which have isotropic correlations equal to the weighting parameter $\lambda$, i.e.\ $E = -\lambda \Ident$. Here $\Ident_4=\operatorname{diag}[1,1,1,1]$ denotes an equal mixture of the four Bell states $\ket{\psi^\pm}=\left(\ket{10}\pm\ket{01}\right)/\sqrt{2}$ and $\ket{\phi^\pm}=\left(\ket{00}\pm\ket{11}\right)/\sqrt{2}$. For $1/3 < \lambda \leq 1$ the Werner states are entangled. By choosing $\lambda = 1/3$ we obtain a separable state, which we denote by $\widetilde\rho_W$. However, it has non-zero discord, which, according to the relation we established earlier, awards the state with a limited suitability for RSP quantified by $\mathcal{D}^2 = \mathcal{F} = \lambda^2 = 1/9$.

Useful resource states for RSP can easily be missed when looking only for entanglement. In fact, the situation is more delicate: Not only does entanglement overlook the capability of certain states for RSP, its verdict can actually be misleading. This deficit can be explicitly illustrated by introducing another type of state, $\rho_B$:
\begin{eqnarray}
\rho_B &=& \frac{1-k}{4} \ket{\psi^+}\bra{\psi^+}+\frac{1+3k}{4}\ket{\psi^-}\bra{\psi^-}\\\nonumber
&+& \frac{1-2t-k}{4} \ket{00}\bra{00}+\frac{1+2t-k}{4}\ket{11}\bra{11},
\label{eqn:Boris}
\end{eqnarray}
where $k$ and $t$ are real parameters specifying the mixture of Bell states and computational basis states. Such a state has a Bloch representation with local vectors $\vec{a} = \vec{b} = t \vec{e}_z$ and correlation tensor $E = -k \Ident$ (isotropic correlations). We set $k = 1/5, t = 2/5$ to obtain a state with non-zero entanglement as measured by a concurrence~\cite{Wootters1998} of $\mathcal{C} = 1/5$ and denote it by $\widetilde\rho_B$. It is tempting to assume that this state might be better suited for RSP than any separable state. However, the RSP-fidelity of $\mathcal{F} = \mathcal{D}^2 = k^2 = 1/25$, compared to the value of $\mathcal{F}=1/9$ for the state $\widetilde\rho_W$, shows that this is not the case. This underlines that entanglement does not qualify as a distinctive resource for RSP. This possibility was also anticipated in the work of Chaves and de Melo~\cite{CM2001}, but here we show that geometric discord can assess the situation correctly.

\begin{figure}
\includegraphics[width=0.4\textwidth]{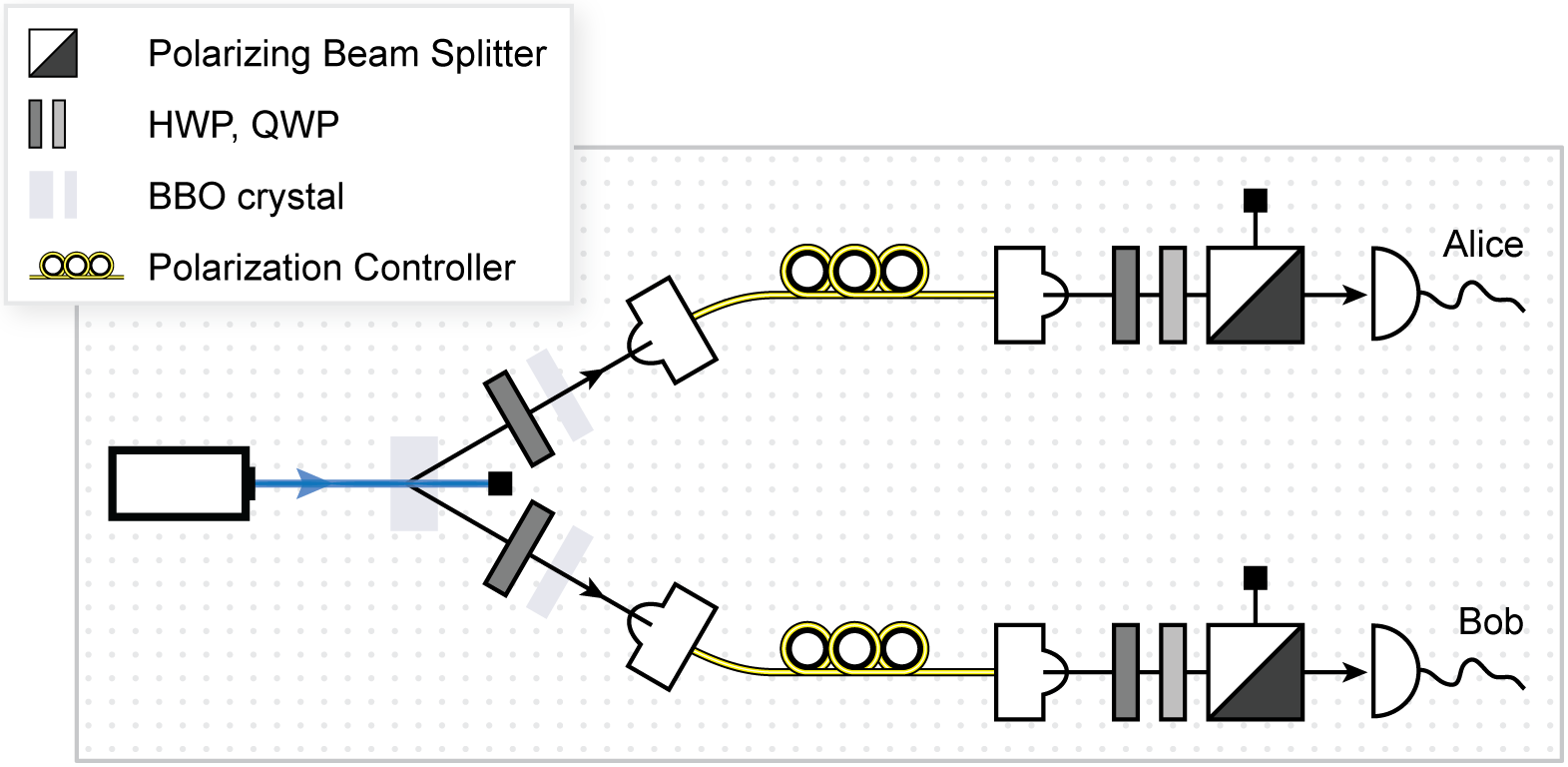}
\label{fig:RSPSetup}
\caption{Experimental setup to realize the RSP protocol. A $\beta$-barium borate (BBO) crystal  is pumped with laser beam (394.5 nm, 76 MHz) such that entangled photon pairs are created via type-II spontaneous parametric down-conversion (SPDC) at a wavelength of 789 nm. The photons are spectrally and spatially filtered by 3 nm narrow-bandwidth filters and by coupling into single-mode fibres, respectively. All four Bell states $\ket{\phi^{\pm}}$, $\ket{\psi^{\pm}}$ can be obtained by rotating polarizations and introducing phase shifts. A combination of quarter-wave plate (QWP), half-wave plate (HWP) and polarizing beam splitter (PBS) is used for the characterization of the prepared quantum states.}
\end{figure}

\textbf{\emph{Experiment and results.--}} We have implemented an experimental test of the RSP protocol using polarization-encoded photonic qubits (Figure~2). We generate all four Bell states $\ket{\phi^\pm}$, $\ket{\psi^\pm}$ and the product states $\ket{00}$, $\ket{11}$ to have access to all the states given in Eq.~\eqref{eqn:Werner} and Eq.~\eqref{eqn:Boris}. For the RSP protocol Alice remotely prepares 58 states uniformly distributed on Bob's Bloch sphere for each of the two resource states~$\widetilde\rho_W$ and~$\widetilde\rho_B$.

\begin{figure*}
\includegraphics[width=15cm]{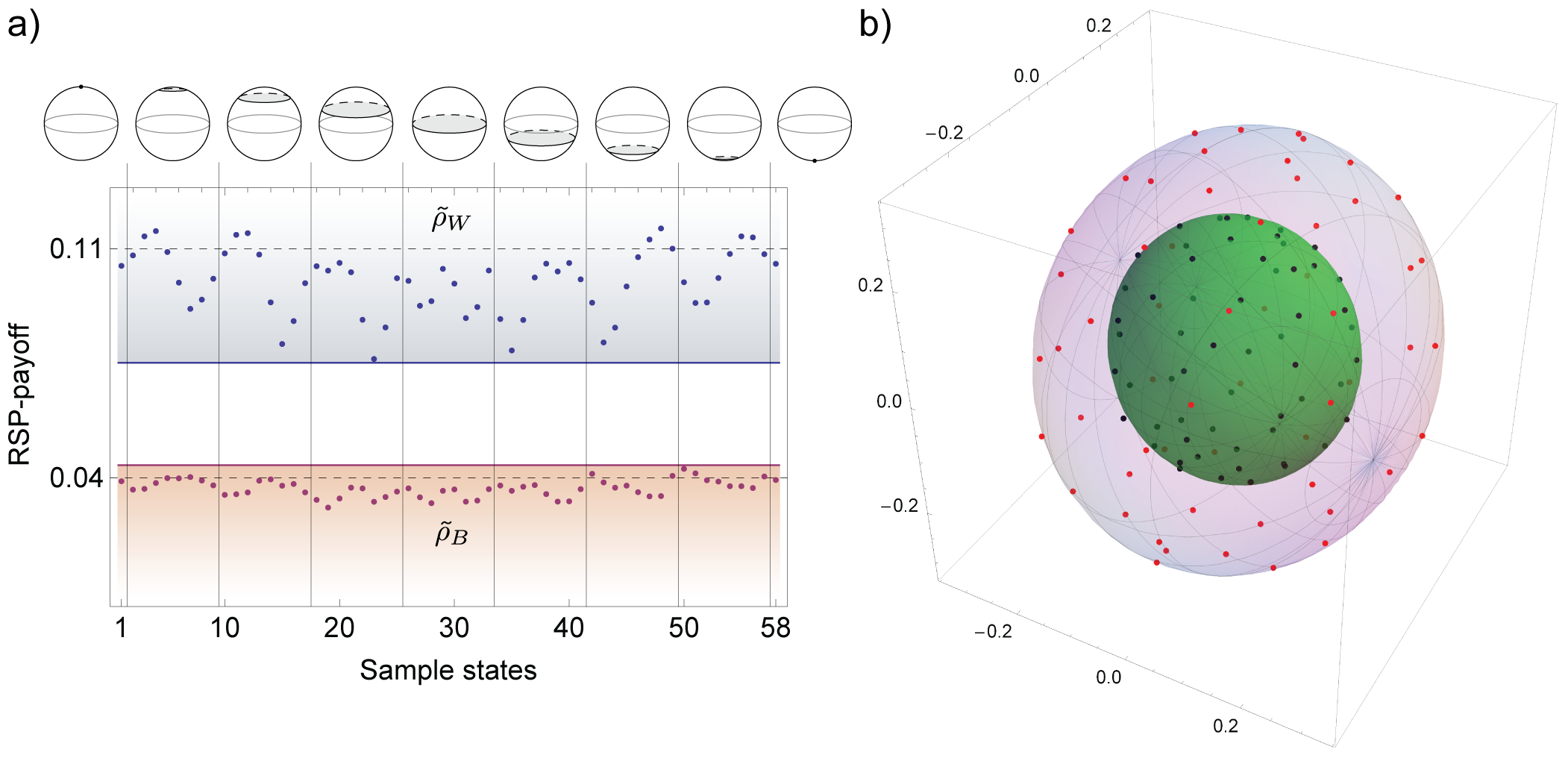}
\caption{Experimentally achieved RSP-payoff $\mathcal{P}$ for 58 distinct states on Bob's Bloch sphere. \textbf{a)} Shown are the respective values for the two resource states $\widetilde\rho_W$ (red) and $\widetilde\rho_B$ (blue). The dashed lines represent the theoretical expectations. There is a clear separation between the two resource states, which indicates that the separable state $\widetilde\rho_W$ is a better resource for RSP than the entangled state $\widetilde\rho_B$. The errors from Poissonian noise are below $6\times 10^{-4}$, which is smaller than the point size. Oscillations in the measured data result from imperfections in the optical setup and demonstrate the sensitivity of the measurements even for mixed quantum states. \textbf{b)} Each state is shown on the Bloch sphere. The outer red points (purple sphere) correspond to $\widetilde\rho_W$ and the inner black points (green sphere) are the same quantum states using $\widetilde\rho_B$ as a resource.}
\label{fig:Payoff}
\end{figure*}

To evaluate the data for the intended resource states we combine the corresponding coincidence counts from different states constituting the mixtures. The relative weights in the mixture are accounted for by appropriate measurement durations. This approach results in state fidelities higher than $0.99$, where the statistical errors, constituting a lower limit, are below~$10^{-4}$ for~$\widetilde\rho_W$  and~$\widetilde\rho_B$.

The characterization of $\widetilde\rho_W$ shows that this state is indeed separable as indicated by a vanishing concurrence and has a higher value of geometric discord than the entangled state $\widetilde\rho_B$ (see Table~1). The remotely prepared states and the respective payoffs $\mathcal{P}$ are presented in Figure~3. We find the separation $\Delta\mathcal{P}=\mathcal{P}_{\widetilde\rho_W} - \mathcal{P}_{\widetilde\rho_B}$ between two corresponding values to be larger than $\Delta\mathcal{P}=0.0434\pm0.0007$ for all prepared states which confirms the better performance of the separable state $\widetilde\rho_W$ by 62 standard deviations. Although the prepared resource states are of high state fidelity, smallest experimental imperfections, in the form of a slight rotation of the Bloch sphere axis, lead to fluctuations in the data (see Figure~3(a)). This effect leads to periodic oscillations instead of the expected constant behaviour of values.

\textbf{\emph{Conclusion.--}} We showed that non-zero quantum discord is the optimal resource for remote state preparation. This is demonstrated by using a variety of polarization-correlated photon pairs. Furthermore, we show that the geometric measure of quantum discord is directly linked to the fidelity of the remote state preparation for a broad class of states, providing an operational interpretation for this measure. Our demonstration that separable states can achieve higher fidelities in remote state preparation than entangled states underlines that not entanglement, but quantum discord quantifies the non-classical correlations required for the task. This insight might be of importance for future quantum-enhanced applications that rely on resources different from quantum entanglement.

\begin{table}
\begin{center}
\begin{tabular}{r|c c}
 & state $\widetilde\rho_W$ & state $\widetilde\rho_B$ \\
\hline
state fidelity & $0.998$ & $0.993$ \\
purity & $0.33$ & $0.36$ \\
concurrence & $0.00$ & $0.12$ \\
geometric discord & $0.097$ & $0.036$ \\
RSP-fidelity & $0.098$ & $0.036$ \\
\end{tabular}
\label{tab:StateData}
\caption{Characterization of the experimentally created states $\widetilde\rho_W$ and $\widetilde\rho_B$. State fidelity, purity and concurrence have been extracted from the density matrices, while the geometric discord has been calculated according to Ref. [15] and the RSP-fidelity using the procedure given in the methods section. The errors computed by simulating Poissonian counting statistics are below $10^{-3}$.}
\end{center}
\end{table}

\section*{Methods}
We show that Alice can achieve non-zero RSP-fidelity if and only if the initial state has non-zero quantum discord, i.e.\ it cannot be represented by Eq.~\eqref{eqn:ClassicalState}. We then show, that for a certain class of states the geometric measure of quantum discord is equivalent to the fidelity of the RSP protocol. Let us first calculate the payoff of Alice in a single run, i.e.\ for a given $\vec s$. By measuring along $\vec \alpha$ she obtains one of the two results $\alpha = \pm 1$ with probability $P(\alpha) = \frac{1}{2}(1+\alpha \, \vec \alpha \cdot \vec a)$. Depending on her result the state of Bob is projected onto
\begin{equation}
\vec b_{\alpha} = \frac{\vec b + \alpha E^T \vec \alpha}{1+\alpha \, \vec \alpha \cdot \vec a},
\label{eqn:collapse}
\end{equation}
where $E$ is the $3 \times 3$ correlation tensor $E_{kl}$ and the superscript $T$ stands for transposition. Without Bob applying the conditional rotation his state is a mixture $P(+1) \vec b_+ + P(-1) \vec b_-=\vec b$. With the rotation, the component of $\vec b_-$ in the plane orthogonal to $\vec \beta$ is flipped and therefore the state $\vec r$ on Bob's side takes the form $\vec r=P(+1) \vec b_+ + P(-1)R_{\pi} \vec b_-$, where $R_{\pi}$ is the rotation that is applied by Bob. To evaluate the payoff-function $\mathcal{P}$ we compute the scalar product $\vec r \cdot \vec s = (P(+1) \vec b_+ - P(-1)\vec b_-) \cdot \vec s$, as $(R_{\pi}\vec b_-)\cdot\vec s = - \vec b_- \cdot\vec s$. Hence $\mathcal{P}$ takes the form
\begin{equation}
\mathcal{P} = \left( \vec \alpha E \vec s \right)^2.
\end{equation}
To find the maximum of $\mathcal{P}$ attainable by Alice's choice of $\vec \alpha$ we introduce a coordinate system in which the direction of $\vec \beta$ corresponds to the $z$-direction of Alice, hence $\vec s = (s_1,s_2,0)$. By expanding the matrix multiplication, the payoff in this frame is given by a scalar product $\mathcal{P} = (\vec \alpha\cdot \vec e)^2$ between the vectors $\vec \alpha$ and $\vec e$ with components $e_j = E_{j1} s_1 + E_{j2} s_2$. Since $\vec \alpha$ is an arbitrary normalized vector of Alice's choice, she can optimize the payoff by choosing it parallel to $\vec e$. The optimized payoff is given by
\begin{equation}
\mathcal{P}_{opt} = \sum_{j=1}^3 (E_{j1} s_1 + E_{j2} s_2)^2.
\label{eqn:Popt}
\end{equation}
Her expected payoff is averaged over the distribution of vectors $\vec s=(\cos\phi,\sin\phi,0)$ (on a circle perpendicular to $\vec \beta$) and reads
\begin{eqnarray}
\langle \mathcal{P}_{opt} \rangle &=& \frac{1}{2\pi}\int_{0}^{2\pi}\mathrm{d}\phi\sum_{j=1}^3 (E_{j1} \cos\phi + E_{j2} \sin\phi)^2 \nonumber\\
 &=&\frac{1}{2} \left( E_{11}^2 +E_{21}^2 + E_{31}^2 + E_{12}^2 + E_{22}^2 + E_{32}^2\right).
\label{eqn:AvgP}
\end{eqnarray}
Since $\vec \beta$ is an arbitrary direction we shall analyze the worst-case scenario to determine the overall fidelity of the RSP protocol for a given resource state. This is achieved by minimizing $\langle \mathcal{P}_{opt} \rangle$ over $\vec \beta$, which in our present notation is the $z$-axis of Alice. Since $\left\|E\right\|^2=\sum_{k,l=1}^3 E_{kl}^2$ is a constant independent of the choice of the local coordinate systems, minimization of the six elements on the right-hand side of Eq.~\eqref{eqn:AvgP} is equivalent to maximization of the sum of the three remaining elements $E_{13}^2 + E_{23}^2 + E_{33}^2$. We can rewrite $E_{13}^2 + E_{23}^2 + E_{33}^2 = \vec{\beta}^T(E^TE)\vec{\beta}$, hence it's maximum is the largest eigenvalue of $E^\mathrm{T}E$. We define the RSP-fidelity as the minimal average payoff
\begin{equation}
\mathcal{F} = \min_{\vec \beta} \langle \mathcal{P}_{opt} \rangle = \frac{1}{2}(E_2^2 + E_3^2) ,
\label{eqn:Fidelity}
\end{equation}
where the eigenvalues of $E^\mathrm{T}E$ are denoted by $E_1^2\geq E_2^2\geq E_3^2$. This quantity captures the quality of RSP for a given resource state. Note that in the case of isotropic correlations all the measures $\mathcal{P}$, $\mathcal{P}_{opt}$, $\langle \mathcal{P}_{opt}\rangle$ and $\mathcal{F}$ coincide.

As we show next, the RSP-fidelity matches exactly the geometric measure of quantum discord for a broad class of states. According to Ref. [15], the (normalized) geometric measure of quantum discord can be written as
\begin{equation}
\mathcal{D}^2 = \frac{1}{2} \left( \|\vec a\|^2 + \|E\|^2 - k_{max} \right) ,
\label{eqn:gdDakic}
\end{equation}
where $\|E\|^2 = \Tr(E^\mathrm{T} E)$, $\vec a$ is the local Bloch vector, $E$ is the correlation tensor and $k_{max}$ is the largest eigenvalue of $K=\vec a\vec a^\mathrm{T} + EE^\mathrm{T}$. Consider now a state where the local Bloch vector $\vec a$ is parallel to the eigenvector corresponding to largest eigenvalue of $E^\mathrm{T} E$. In the eigenbasis of the correlation tensor~\cite{Horodecki1996} the matrix $E^\mathrm{T} E$ has diagonal form $E^\mathrm{T} E =\operatorname{diag}[E_1^2,E_2^2,E_3^2]$, where $E_1^2 \geq E_2^2 \geq E_3^2$ are the eigenvalues of $E^\mathrm{T} E$. In this basis the local Bloch vector is of form $\vec a = (\kappa,0,0)$ and therefore the matrix $K$ is also diagonal $K=\operatorname{diag}[E_1^2+\kappa^2,E_2^2,E_3^2]$ with largest eigenvalue $k_{max}=E_1^2+\kappa^2$. Hence Eq.~\eqref{eqn:gdDakic} simplifies to
\begin{align}
\mathcal{D}^2 =& \frac{1}{2} \left( \|\vec a\|^2 + \|E\|^2 - k_{max} \right) \nonumber\\
=& \frac{1}{2} \left( |\kappa|^2 + E_1^2+E_2^2+E_3^2 - (E_1^2+\kappa^2) \right) \nonumber\\
=& \frac 1 2 \left( E_2^2+E_3^2 \right) .
\label{eqn:gdDakicSimplify}
\end{align}
This result in particular holds for states with maximally mixed marginals ($\vec a = \vec b = 0$), where $\kappa=0$. Further examples of states in the described class are all isotropically correlated states ($E=\lambda \Ident$), as there are no restrictions on the local Bloch vector $\vec a$ in this case. Therefore the experimentally tested states $\rho_B$ and $\rho_W$ fall into this category.

\textbf{\emph{Acknowledgements.--}}
We acknowledge support from the European Commission, Q-ESSENCE (No 248095), ERC Advanced Senior Grant (QIT4QAD) and the ERA-Net CHIST-ERA project QUASAR, John Templeton Foundation, Austrian Science Fund (FWF): [SFB-FOCUS] and [Y585-N20] and the doctoral programme CoQuS, and the Air Force Office of Scientific Research, Air Force Material Command, USAF, under grant number FA8655-11-1-3004. The work is supported by the National Research Foundation and Ministry of Education in Singapore.


\bibliography{RSPrefs}
\bibliographystyle{naturemag}



\end{document}